Heuristic model on the origin of the homochirality of life

Vladimir Subbotin[1,2*] and Gennady Fiksel[2]

[1]Arrowhead Pharmaceuticals, Madison, WI, USA; [2]University of Wisconsin, Madison, WI, USA

**Abstract**

Life demonstrates remarkable homochirality of its major building blocks: nucleic acids, amino acids, sugars, and phospholipids. We propose a mechanism that places the root of life's homochirality in the formation of phospholipid bilayer vesicles (liposomes). These liposomes are formed at the water/air interface from Langmuir layers and contain ribose, presumably delivered to Early Earth by carbonaceous meteorites. Although the extraterrestrial ribose was initially racemic, life is homochiral, based on D-ribose and its derivatives. The phospholipid membrane's high permeability to D-ribose, combined with the ribose's interaction with the bilayer's charged phosphate groups, leads to ribose phosphorylation, forming D-ribose-5-phosphate. Once inside, the D-ribose-5-phosphate molecules cannot cross the membrane. The catalytic action of $Fe^{3+}$ ions greatly enhances the phosphorylation rate. Overall, this process is enantioselective, substantially favoring the buildup of D-ribose over L-ribose. Through liposome fusion, fission, and self-replication, this eventually leads to the Darwinian evolution of these structures and to the conversion of D-ribose-5-phosphate into complex functional molecules, such as ribozymes and RNA, and eventually into DNA, all of which inherit D-ribose's chirality.

**Introduction**

Several scenarios of the origin of life, e.g., RNA World and Metabolism First, assume the occurrence of multiple events that must align strictly consecutively in time and space, i.e., appear as a multistage process. However, such staging raises strong skepticism, for example, by A.I. Oparin [1] and L.E. Orgel [2]. An alternative model, Lipid First [3,4], places the origin of life into self-assembling lipid molecules and the formation of liposomes before more complex, information-carrying RNA emerged. This process, through self-replication, fusion, and fission, eventually leads to Darwinian evolution of these simple, spontaneously organized structures.

We have further advanced this concept by proposing a new hypothesis of self-sustaining Darwinian evolution of liposomes [5] that relies solely on natural and ubiquitous phenomena: solar UV radiation, the day/night cycle, and gravity. In our scenario, liposomes formed at the Langmuir layer would be inevitably destroyed by solar UV unless they acquire negative buoyancy by encapsulating heavy solutes, such as ribose, and descend from the water/air interface. We have also shown that some primordial water constituents, for example, ferric salts, can provide UV attenuation strong enough to protect liposomes at depths of several millimeters [6,7]. The process forms resilient, UV-protected, autocatalytic membranes; ensures liposomal survival, fusion, fission, and component mutation; and provides liposomes' adaptation and persistence. The new hypothesis comprises the necessary prerequisites for Darwinian evolution: mutable adaptive traits, heredity, and selective forces. A pictorial representation of the new hypothesis is depicted in Figure 1.

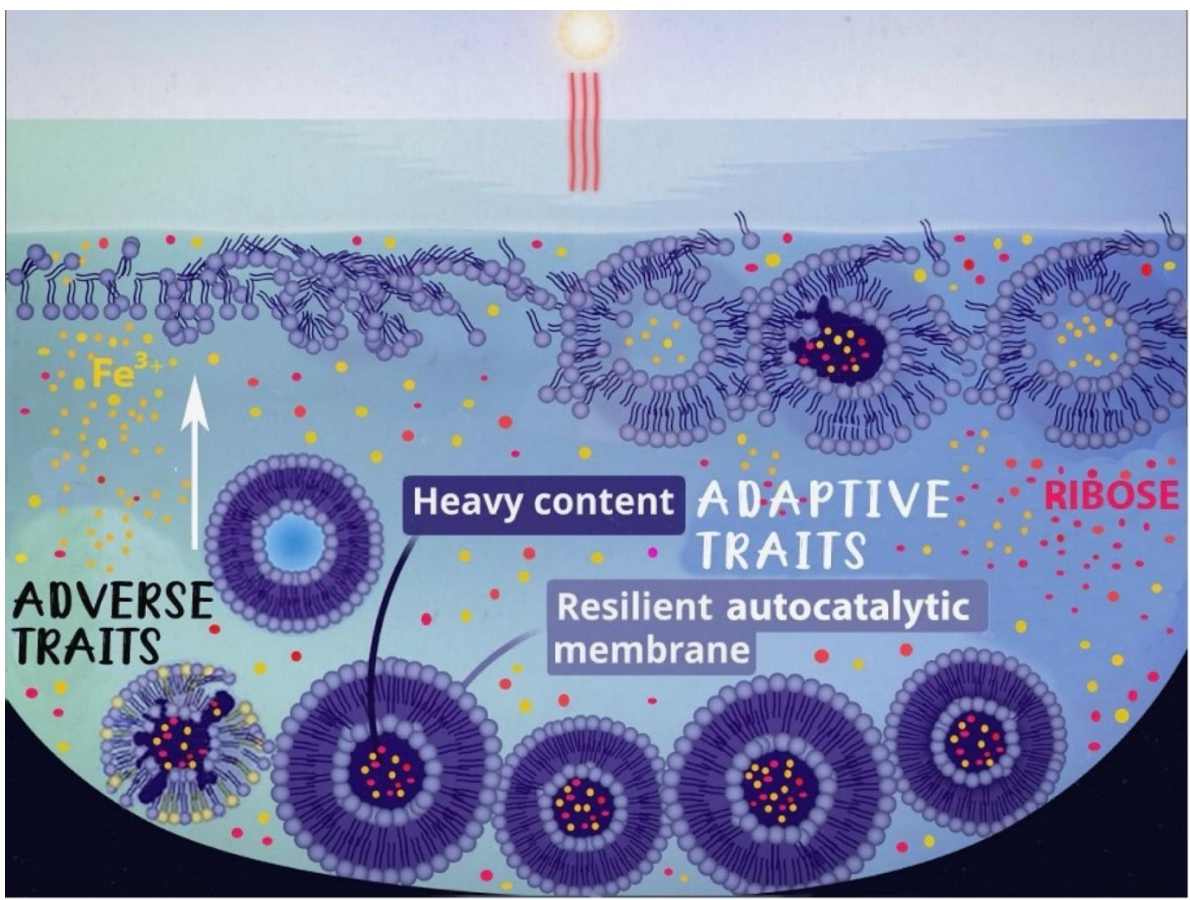

**Figure 1**. Artistic depiction of liposome formation and evolution. Liposomes formed on the surface of the Archean water pools are destroyed by Solar UV unless they acquire two adaptive traits – the heavy content that sinks the liposomes to the bottom of the pool and facilitates protection from UV, and the formation of resilient autocatalytic membrane composition that ensures liposomal survival, fusion, fission, and mutation of the components, thus providing liposomes adaptation and persistence.

**A model of the origin of life's homochirality**

Historically, the issue of life's homochirality has been explored through various concepts, most notably "The Frozen Accident Theory " by Francis Crick, published in 1968 [8] and later extended by others [9-12]. Additionally, other concepts were considered, including *ab initio* homochirality [13] as well as cosmological phenomena of different natures and magnitudes: supernovae explosions [14,15], weak nuclear force [16], asymmetric dark matter [17], circularly polarized light generated in the interstellar medium [18], and the concepts of the "sterile" and the left-handed neutrinos [19]. Despite these efforts, homochirality remains a great mystery, perhaps no less perplexing than the Origin of Life itself [19-22].

We propose an alternative mechanism connecting the origin of life's homochirality to the formation of phospholipid bilayer vesicles (liposomes) and to specificities in the iterative evolution of an initially racemic mixture of D- and L-ribose [23].

Ribose delivered by carbonaceous meteorites is one of the most likely heavy cargo candidates providing the submersion of the liposomes. Meteorite-delivered ribose appeared as racemic, meaning it contained equal parts of D-ribose and L-ribose [24-26]. However, modern life consists only of D-ribose and its derivatives. Numerous studies, theoretical and experimental [27-31], have shown that, due to the high rate of autocatalytic reactions, even a slight imbalance in the initial concentrations of the two enantiomeric forms of molecules can be amplified over time. But this overlooks situations in which a slight initial preference for one enantiomeric form, such as D-ribose at one site, can be counterbalanced by the opposite preference at a nearby site. Additionally, despite the high rate of autocatalytic reactions, the outcome still would strongly depend on the initial concentration disparity - the greater the initial difference, the more likely it is to influence the final result. Clearly, there is a need for a strong, robust, and universal mechanism favoring the prevalence of D-ribose, but it hasn't been firmly established yet.

Biological membranes, and particularly phospholipid membranes, are enantioselectively permeable [32-36]. In particular, recent results by Goode et al. [37] demonstrated that Archaeal and hybrid membranes (all containing phospholipids) exhibit significantly greater permeability to D-ribose than to L-ribose. However, the high D-ribose permeability alone *does* not solve the puzzle: the same amount of D-ribose that enters the compartment must also exit it, meaning that in equilibrium, an initially racemic mixture would remain racemic.

**Key chemical assumption and its experimental status**

The mechanism we suggest involves the **selective accumulation of phosphorylated D-ribose within phospholipid membranes**, breaking the racemic equilibrium and enabling the next evolutionary step: converting trapped D-ribose-5-phosphate into complex functional molecules, such as ribozymes, RNA, and ultimately DNA.

We assume that on early Earth, pathways already existed that converted simple lipid membranes into phospholipids [38-40]. A phospholipid membrane can donate a phosphate group ($PO_4^{3-}$) to a D-ribose passing through the membrane via a transphosphorylation reaction. In most circumstances, the transphosphorylation reaction is slow because it requires breaking strong P-O bonds in the phospholipid. However, in the presence of mineral catalysts, such as $Fe^{3+}$, the phosphorylation rate can be increased by orders of magnitude, even in aqueous environment [41-42]. In addition, it was established [43] that the rate of D-ribose binding to phosphate is higher than for other sugars.

A phosphorylated, negatively charged D-ribose might then become electrostatically trapped within the membrane. Other mechanisms limiting membrane crossing are also possible. Phosphorylation-reduced membrane-crossing behavior was studied in [44], and a requirement for a specialized transporter in plants' membranes was demonstrated in [45]. All these selectively accumulate D-ribose inside the liposome and increase its concentration faster than L-ribose. A simplified artistic illustration of the model is depicted in Figure 2.

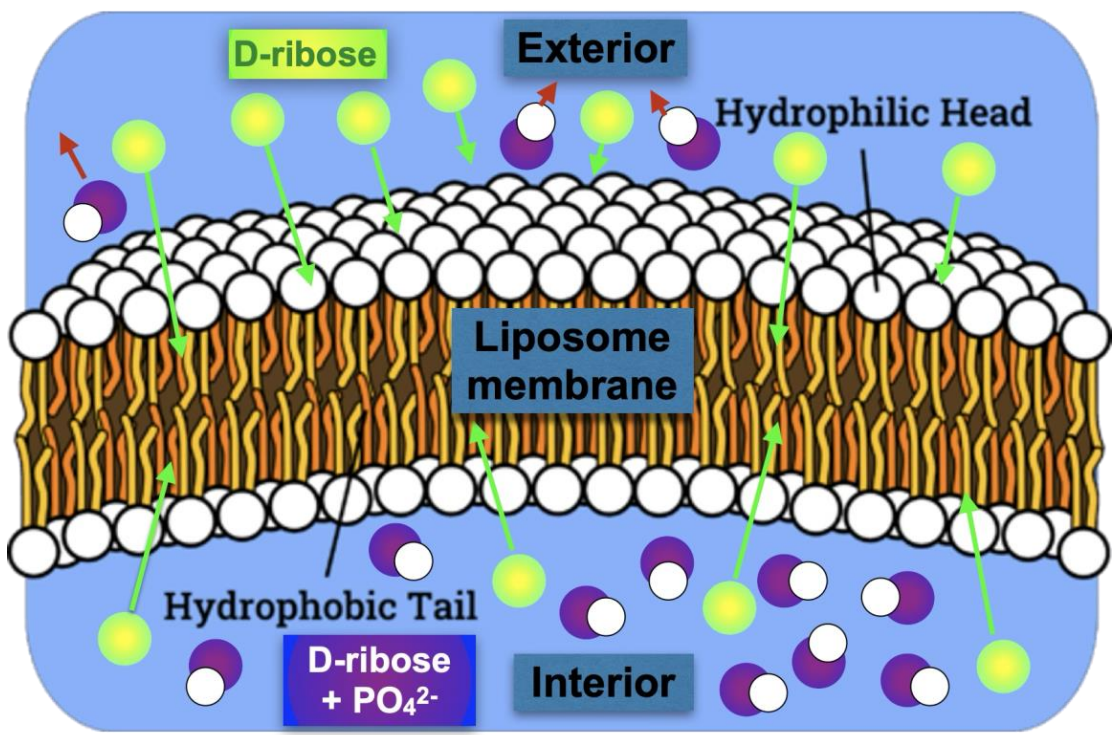

**Figure 2.** A simplified artistic illustration of the model. It depicts a cross-section of a spherical phospholipid vesicle, highlighting its membrane with hydrophilic heads and hydrophobic tails. Racemic D- and L-ribose (green) can penetrate the membrane, but only D-ribose goes through phosphorylation and becomes trapped (purple) within the membrane.

The proposed concept - (a) self-assembly of phospholipid liposomes in Archean waters containing a mixture of D- and L-ribose, (b) liposome submergence that provides protection against solar UV and at the same time is greatly enhanced by the presence of $Fe^{3+}$ in Fe-rich Archean waters, and (c) the selective accumulation of phosphorylated D-ribose, also boosted by the presence of ferric salts, collectively creates a comprehensive and self-contained framework describing the origin of modern life's chirality. Notably, the same mineral, Fe, essential for UV protection, is also vital for initiating abiogenetic evolution!

**Simple mathematical model**

The process outlined in the previous section can be modeled by simple balance equations. Consider the following assumptions:

1. A spherical liposome with a radius $R$ containing a racemic mixture of D- and L-ribose at equal concentrations is immersed in a large prebiotic water pool containing the same racemic mixture at the same concentrations.
2. Ribose attaches a negatively charged phosphate group while passing through the membrane (phosphorylation) and becomes trapped inside the membrane by electrostatic forces.

In a simple case of symmetric membrane permeability $P$, the total ribose flux per unit area through the membrane is $I = (n_{out} - n_{in})P$, where $n_{out}$ and $n_{in}$ are the ribose densities outside and inside the membrane, respectively. Clearly, in the absence of any particle source, the equilibrium condition $I = 0$ requires equal densities $n_{in} = n_{out}$ on both sides, regardless of permeability. Therefore, if the initial ribose outside is racemic, it will remain racemic inside the membrane as well.

Now, include the process of capturing the phosphate group in the membrane. If during its passage through the membrane, a ribose undergoes phosphorylation at a volumetric reaction rate $K_p$, then the probability of the reaction is $1 - \exp(-K_p \delta / P)$, where $\delta$ is the membrane thickness. For a small $\delta$, such that $K_p \delta / P \ll 1$, the fraction of the flux that undergoes phosphorylation is $K_p \delta / P$. To verify this assumption, substitute $K_p = 10^{-5}$ s$^{-1}$ (a high value, but not unreasonable in the presence of Fe$^{3+}$ catalyst, $\delta = 5 \times 10^{-4}$ cm, and P = $50 \times 10^{-8}$ cm/s [46], which would yield $K_p \delta / P = 0.01$.

The behavior of non-phosphorylated density $n_{in}$ within the membrane can be described by

$$V \frac{dn_{in}}{dt} = A n_{out} P \left(1 - \frac{K_p \delta}{P}\right) - A n_{in} P, \#(1)$$

where $V = 4/3 \pi R^3$ is the vesicle volume, and $A = 4\pi R^2$ is its surface area. The first term on the RHS of Eq. (1) represents the inward flux adjusted by the volumetric phosphorylation. The second term is the outward flux. Clearly, under the assumption of $K_p \delta/P \ll 1$, the equilibrium inside density is very close to that outside $n_{in} \approx n_{out}$.

Assume further, for simplicity, a perfect trapping of the phosphorylated D-ribose. Later, we expand on this assumption. The phosphorylated part of the inward flux determines the behavior of the trapped phosphorylated density within the membrane,

$$V \frac{dn_p}{dt} = A n_{out} K_p \delta. \#(2)$$

Note here that not all the inward flux reaches the membrane's interior. Some of it, after phosphorylation near the outer boundary, would be repelled outward by the electrostatic forces. However, by symmetry, this loss would be exactly balanced by non-phosphorylated outward flux that is converted near the inner boundary and repelled inward.

Considering all the points mentioned, the solution for the trapped phosphorylated ribose density is

$$n_p(t) = n_{out}(1 + t/\tau_d), \#(3)$$

where the characteristic "doubling" time τ is

$$\tau_d = \frac{V}{AK_p\delta} = \frac{1}{3K_p}\frac{R}{\delta} . \#(4)$$

As a numerical estimate, assume a typical shell geometry of $R = 50\ \mu m$, $\delta = 5\ \mu m$, and the volumetric reaction rate $K_p = 10^{-5}\ s^{-1}$ resulting in a characteristic time $\tau_d = 92$ hours. Therefore, after just under four days, the concentration of D-ribose inside the membrane will **double** compared to that of L-ribose, which was presumed not to undergo phosphorylation-mediated accumulation.

Add a loss term for the phosphorylated D-ribose and modify Eq. (2) accordingly.

$$\frac{dn_p}{dt} = \frac{n_{out}}{\tau_d} - \frac{n_p}{\tau_{conf}}, \#(5)$$

where $\tau_d$ is defined by Eq. (4) and $\tau_{conf}$ is a confinement time of a generic loss mechanism, combining all possible losses, for example, the membrane leakage, the ribose disassembly, etc. If each of these losses can be characterized by its own confinement times $\tau_{conf1}, \tau_{conf2} \ldots$ , then the overall confinement is determined by $1/\tau_{conf} = 1/\tau_{conf1} + 1/\tau_{conf2} \ldots$ .

Note that the equilibrium phosphorylated D-ribose density is $n_p = n_{out}\tau_{conf}/\tau_d$, so the higher the ratio $\tau_{conf}/\tau_d$, the higher the final D-ribose density. A time-dependent solution of Eq. (5) is:

$$\frac{n_p}{n_{out}} = \frac{\tau_{conf}}{\tau_d} - \left(\frac{\tau_{conf}}{\tau_d} - 1\right)e^{-t/\tau_{conf}}. \#(6)$$

Figure 3 illustrates the time evolution of the D-ribose buildup factor $n_p/n_{out}$. Each curve asymptotically approaches its equilibrium value $\tau_{conf}/\tau_d$, with the growth rate determined by the phosphorylation rate $K_p$.

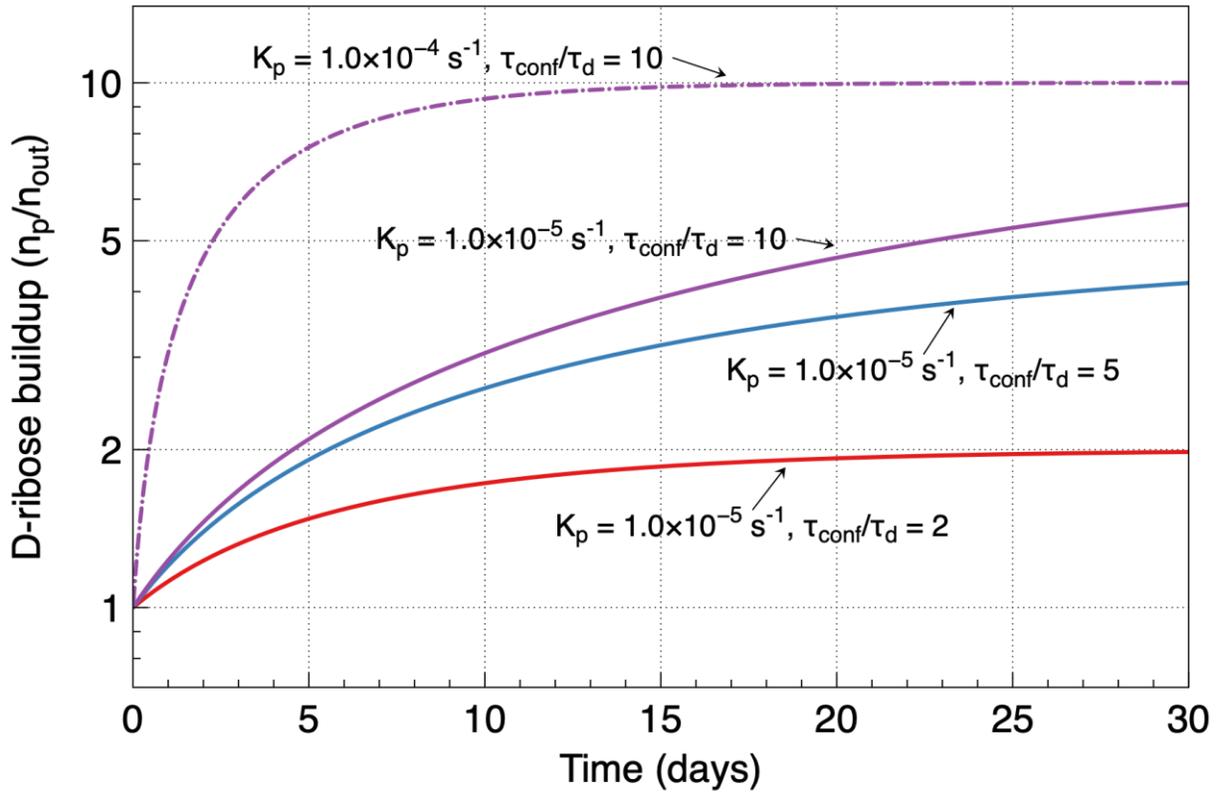

**Figure 3**. A time evolution of the D-ribose buildup factor $n_p/n_{out}$. Solid lines are plotted for a phosphorylation rate $K_p = 10^{-5}$ s$^{-1}$ and different ratios $\tau_{conf}/\tau_d$. The dashed line illustrates the time evolution at a faster phosphorylation $K_p = 10^{-4}$ s$^{-1}$.

**Model discussion**

The above model clearly does not cover all potential scenarios or include every relevant physical and chemical detail. Instead, we intentionally simplified the problem to emphasize the key aspects, aiming to showcase a robust, promising, and self-contained scenario while offering an approximate estimate and the bounds of the effect.

Undoubtedly, a significant impact would come from the proposed stereoselectivity of D-ribose phosphorylation, especially in the presence of $Fe^{3+}$. In the model, we completely ignored L-ribose phosphorylation. As mentioned earlier, some studies [42,43,47,48] have suggested that the D-ribose enantiomer generally has a higher phosphorylation rate than L-ribose, in particular due to the specific stereochemical interactions between D-ribose and Fe ions. Whether this is accurate relative to our prebiotic scenario and to what extent remains unknown. However, because the buildup rate is generally proportional to the phosphorylation rate, even a modest factor (2–5×) would be sufficient to drive enrichment.

Another factor could be a process that disrupts the linear buildup of D-ribose as described in Eq. (3), such as imperfect trapping or other mechanisms. We estimated this effect by introducing a generalized loss mechanism and demonstrated that the D-ribose buildup factor is proportional to the ratio of the mechanism's confinement time to the phosphorylation time. The latter also determines the growth rate of D-ribose accumulation.

These and other issues could be addressed through experimentation.

**Suggested experimental tests**

Our hypothesis outlined above can be tested in the following experiments:

1. Liposomes with different membrane compositions are loaded with a racemic D/L ribose mixture. The key idea is that different membrane compositions will provide diverse tools for enantiomeric ribose separation [34,49]
2. Liposomes are suspended in a racemic ribose mixture
3. Add varying concentrations of $FeCl_3$.

4. The membrane composition that encourages D-ribose-5-phosphate accumulation in liposomes, leading to D's predominance over L, would be an ideal candidate for further study into biogenesis and the origins of life.

**Conclusions**

We propose a mechanism connecting the origin of life's homochirality to the formation of phospholipid bilayer vesicles (liposomes) and to specificities in the iterative evolution of an initially racemic mixture of D- and L-ribose. D-ribose, passing through the membrane, can bind to the membrane's phosphate group and become trapped inside. The presence of $Fe^{3+}$, a catalytic agent, could significantly increase the phosphorylation rate and preferentially enhance the efficiency of D-ribose accumulation over L-ribose. Our model showed that, in just a few days, the concentration of D-ribose inside the membrane vesicle would double. We discuss key factors, such as phosphorylation rate and D-ribose losses. We acknowledge that some supporting experiments have been conducted in studies such as mineral-assisted phosphorylation or the permeability of plants' membranes, rather than in the prebiotic environment we are describing. We hope the proposed experimental tests will shed further light on these critical issues.

The selective accumulation of phosphorylated D-ribose leads to the formation of D-deoxyribose-5-phosphate. Both D-ribose and D-deoxyribose are crucial for establishing the chirality of D-RNA and D-DNA. Additionally, the homochirality of D-deoxyribose/D-ribose nucleic acids determines the L-isoform of amino acids and proteins [15,50-54].

In summary, the proposed hypothesis organically integrates concepts proposed earlier: (a) the self-assembly of phospholipid liposomes in Archaean waters containing a mixture of D- and L-ribose, (b) liposome submergence that provides protection against solar UV, and (c) selective accumulation of phosphorylated D-ribose. The presence of $Fe^{3+}$ boosts both UV protection and selective accumulation of phosphorylated D-ribose. Collectively, these elements form a comprehensive, self-contained framework that explains the origin of the chirality of modern life.

*Address correspondence to:
Vladimir Subbotin, Arrowhead Pharmaceuticals, Madison, WI 53719, USA
vsubbotin@arrowheadpharma.com

**References**

**1** Oparin, A. (1974) A hypothetic scheme for evolution of probionts. *Origins of life* 5 (1), 223-226

**2** Orgel, L.E. (1998) The origin of life—a review of facts and speculations. *Trends in biochemical sciences* 23 (12), 491-495

**3** Hargreaves, W. et al. (1977) Synthesis of phospholipids and membranes in prebiotic conditions. *Nature* 266 (5597), 78-80

**4** Luisi, P.L. et al. (2004) A possible route to prebiotic vesicle reproduction. *Artificial Life* 10 (3), 297-308

**5** Subbotin, V. and Fiksel, G. (2023) Exploring the lipid world hypothesis: a novel scenario of self-sustained darwinian evolution of the liposomes. *Astrobiology* 23 (3), 344-357

**6** Subbotin, V. and Fiksel, G. (2023) Aquatic ferrous solutions of prebiotic mineral salts as strong UV protectants and possible loci of life origin. *Astrobiology* 23 (7), 741-745

**7** Turner, B. et al. (2025) Protection of liposomes by ferric salts against the UV damage and its implications for the origin of life. *Frontiers in Astronomy and Space Sciences* 12, 1566396

**8** Crick, F.H. (1968) The origin of the genetic code. *Journal of molecular biology* 38 (3), 367-379

**9** Root-Bernstein, R. (2010) Experimental test of L-and D-amino acid binding to L-and D-codons suggests that homochirality and codon directionality emerged with the genetic code. *Symmetry* 2 (2), 1180-1200

**10** Lacey Jr, J.C. and Mullins Jr, D.W. (1985) Genetic coding catalysis. *J Theor Biol* 115 (4), 595-601

**11** Weber, A.L. and Miller, S.L. (1981) Reasons for the occurrence of the twenty coded protein amino acids. *Journal of molecular evolution* 17 (5), 273-284

**12** Root-Bernstein, R. (2007) Simultaneous origin of homochirality, the genetic code and its directionality. *Bioessays* 29 (7), 689-698

**13** Pross, A. (2016) *What is life?: How chemistry becomes biology*, Oxford University Press

**14** Bonner, W.A. (1995) Chirality and life. *Origins of Life and Evolution of the Biosphere* 25 (1), 175-190

**15** Cline, D.B. (2005) On the physical origin of the homochirality of life. *European Review* 13 (S2), 49-59

**16** MacDermott, A. and Tranter, G. (1995) SETH: the search for extra-terrestrial homochirality. *Journal of biological physics* 20 (1), 77-81

**17** Yin, W. et al. (2025) Asymmetric warm dark matter: from cosmological asymmetry to chirality of life. *Journal of Cosmology and Astroparticle Physics* 2025 (02), 063

**18** Myrgorodska, I. et al. (2017) Light on chirality: absolute asymmetric formation of chiral molecules relevant in prebiotic evolution. *ChemPlusChem* 82 (1), 74-87

**19** Devínsky, F. (2021) Chirality and the origin of life. *Symmetry* 13 (12), 2277

**20** Chieffo, C. et al. (2023) The origin and early evolution of life: Homochirality emergence in prebiotic environments. *Astrobiology* 23 (12), 1368-1382

**21** Weller, M.G. (2024) The mystery of homochirality on earth. *Life* 14 (3), 341

**22** Budin, I. and Szostak, J.W. (2010) Expanding roles for diverse physical phenomena during the origin of life. *Annual review of biophysics* 39 (1), 245-263

**23** Subbotin, V. and Fiksel, G. (2025) Heuristic model on the origin of the homochirality of life. *arXiv preprint arXiv:2512.11136*

**24** Paschek, K. et al. (2022) Possible ribose synthesis in carbonaceous planetesimals. *Life* 12 (3), 404

**25** Ono, C. et al. (2024) Abiotic ribose synthesis under aqueous environments with various chemical conditions. *Astrobiology* 24 (5), 489-497 26  Abe, S. et al. (2024) Gamma-ray-induced synthesis of sugars in meteorite parent bodies. *ACS Earth and Space Chemistry* 8 (9), 1737-1744
27  Frank, F.C. (1953) On spontaneous asymmetric synthesis. *Biochimica Et Biophysica Acta* 11, 459-463
28  Soai, K. et al. (1995) Asymmetric autocatalysis and amplification of enantiomeric excess of a chiral molecule. *Nature* 378 (6559), 767-768
29  Jeilani, Y.A. and Nguyen, M.T. (2020) Autocatalysis in formose reaction and formation of RNA nucleosides. *The Journal of Physical Chemistry B* 124 (50), 11324-11336
30  Tran, Q.P. et al. (2023) Towards a prebiotic chemoton–nucleotide precursor synthesis driven by the autocatalytic formose reaction. *Chemical Science* 14 (35), 9589-9599
31  Williamson, M.P. (2024) Autocatalytic Selection as a Driver for the Origin of Life. *Life* 14 (5), 590
32  Martin, H.S. et al. (2021) Probing the role of chirality in phospholipid membranes. *ChemBioChem* 22 (22), 3148-3157
33  Hu, J. et al. (2021) Chiral lipid bilayers are enantioselectively permeable. *Nature chemistry* 13 (8), 786-791
34  Hanashima, S. et al. (2020) Enantiomers of phospholipids and cholesterol: a key to decipher lipid-lipid interplay in membrane. *Chirality* 32 (3), 282-298
35  Bocková, J. et al. (2024) The astrochemical evolutionary traits of phospholipid membrane homochirality. *Nature Reviews Chemistry* 8 (9), 652-664
36  Sojo, V. (2015) On the biogenic origins of homochirality. *Origins of Life and Evolution of Biospheres* 45 (1), 219-224
37  Goode, O. et al. (2025) Permeability selection of biologically relevant membranes matches the stereochemistry of life on Earth. *PLoS Biology* 23 (5), e3003155
38  Gull, M. (2014) Prebiotic phosphorylation reactions on the early Earth. *Challenges* 5 (2), 193-212
39  Liu, L. et al. (2020) Enzyme-free synthesis of natural phospholipids in water. *Nature chemistry*, 1-6
40  Zhao, J. and Mimura, K. (2025) Supply of phospholipid precursors and evolution sites on the early Earth by impact. *Geochimica et Cosmochimica Acta*
41  Pinna, S. et al. (2022) A prebiotic basis for ATP as the universal energy currency. *PLoS Biology* 20 (10), e3001437
42  Nam, I. et al. (2017) Abiotic production of sugar phosphates and uridine ribonucleoside in aqueous microdroplets. *Proceedings of the National Academy of Sciences* 114 (47), 12396-12400
43  Cruz, H.A. and Krishnamurthy, R. (2025) Selection of Ribofuranose-Isomer Among Pentoses by Phosphorylation with Diamidophosphate. *Angewandte Chemie International Edition* 64 (35), e202509810
44  Ehringer, W.D. et al. (2000) Membrane permeability of fructose-1, 6-diphosphate in lipid vesicles and endothelial cells. *Molecular and Cellular Biochemistry* 210 (1), 35-45
45  Eicks, M. et al. (2002) The plastidic pentose phosphate translocator represents a link between the cytosolic and the plastidic pentose phosphate pathways in plants. *Plant Physiology* 128 (2), 512-522
46  Santos, T.C. and Futerman, A.H. (2023) The fats of the matter: lipids in prebiotic chemistry and in origin of life studies. *Progress in Lipid Research* 92, 101253
47  Hirakawa, Y. et al. (2022) Borate-guided ribose phosphorylation for prebiotic nucleotide synthesis. *Scientific reports* 12 (1), 11828
48  Banfalvi, G. (2021) Prebiotic pathway from ribose to RNA formation. *International journal of molecular sciences* 22 (8), 3857
49  Fernandes, C. et al. (2017) Chiral separation in preparative scale: A brief overview of membranes as tools for enantiomeric separation. *Symmetry* 9 (10), 206

| | |
|---|---|
| **50** | Fiore, M. (2025) Homochirality emergence: a scientific enigma with profound implications in origins of life studies. *Symmetry* 17 (3), 473 |
| **51** | Banfalvi, G. (2020) Ribose selected as precursor to life. *DNA and Cell Biology* 39 (2), 177-186 |
| **52** | Macdermott, A.J. (1995) Electro weak enantioselection and the origin of life. *Origins of Life and Evolution of the Biosphere* 25 (1), 191-199 |
| **53** | Ikehara, K. (2023) Why Were [GADV]-amino Acids and GNC Codons Selected and How Was GNC Primeval Genetic Code Established? *Genes* 14 (2), 375 |
| **54** | Munegumi, T. (2015) Aldolase as a chirality intersection of L-amino acids and D-sugars. *Origins of Life and Evolution of Biospheres* 45 (1), 173-182 |